\documentclass[journal,a4paper,twoside]{IEEEtran}

\usepackage[]{amsmath}
\usepackage{mathrsfs}
\usepackage{dcolumn}
\usepackage{bm}

\usepackage{graphicx}
 \usepackage[font=footnotesize, caption=false]{subfig}

\usepackage{flushend}

\makeatletter
\long\def\@makecaption#1#2{\ifx\@captype\@IEEEtablestring%
\footnotesize\begin{center}{\normalfont\footnotesize #1}\\
{\normalfont\footnotesize\scshape #2}\end{center}%
\@IEEEtablecaptionsepspace
\else
\@IEEEfigurecaptionsepspace
\setbox\@tempboxa\hbox{\normalfont\footnotesize {#1.}~~ #2}%
\ifdim \wd\@tempboxa >\hsize%
\setbox\@tempboxa\hbox{\normalfont\footnotesize {#1.}~~ }%
\parbox[t]{\hsize}{\normalfont\footnotesize \noindent\unhbox\@tempboxa#2}%
\else
\hbox to\hsize{\normalfont\footnotesize\hfil\box\@tempboxa\hfil}\fi\fi}
\makeatother

\makeatletter
\def\ps@headings{%
\def\@oddhead{\mbox{}\scriptsize\rightmark \hfil }%
\def\@evenhead{\scriptsize \hfil \leftmark\mbox{}}%
\def\@oddfoot{}%
\def\@evenfoot{}}
\makeatother

\makeatletter
\def\ps@IEEEtitlepagestyle{%
\def\@oddhead{\mbox{}\scriptsize\rightmark \hfil }%
\def\@evenhead{\scriptsize \hfil \leftmark\mbox{}}%
\def\@oddfoot{}%
\def\@evenfoot{}}
\makeatother

\begin{document}

\title{Transition Edge Sensor Thermometry for On-chip Materials Characterization}

%

\author{\IEEEauthorblockN{D. J. Goldie\IEEEauthorrefmark{1}, D. M. Glowacka\IEEEauthorrefmark{1},
K. Rostem\IEEEauthorrefmark{1}\IEEEauthorrefmark{2} and S. Withington\IEEEauthorrefmark{1}}\\
\IEEEauthorblockA{\IEEEauthorrefmark{1}Detector and Optical Physics Group\\
Cavendish Laboratory\\ University of Cambridge\\
Madingley Road\\ Cambridge\\CB3 0HE, UK \\
Email: d.j.goldie@mrao.cam.ac.uk}\\
\IEEEauthorblockA{\IEEEauthorrefmark{2}Current Address\\
NASA-Goddard Space Flight Center\\
Greenbelt\\ MD 20771 USA }
}

\markboth{21st International Symposium on Space Terahertz Technology, Oxford, 23-25 March, 2010.}%
{21st International Symposium on Space Terahertz Technology, Oxford, 23-25 March, 2010.}

\maketitle

\begin{abstract}
\boldmath 
The next generation of ultra-low-noise 
cryogenic detectors for space science applications
require continued exploration of materials characteristics at low temperatures.
The low noise and good energy sensitivity of current Transition Edge Sensors (TESs) permits
measurements of thermal parameters of mesoscopic systems 
with unprecedented precision. 
We describe a radiometric technique for differential measurements of 
materials characteristics at low temperatures (below about $3\,{\rm K}$).
The technique relies on the  very  broadband thermal radiation that couples between
impedance-matched resistors that terminate a Nb superconducting microstrip and the power exchanged is
measured using a TES. 
The capability of the TES 
to deliver fast, time-resolved thermometry further expands the parameter space: for example to investigate
time-dependent heat capacity. 
Thermal properties of isolated structures 
can be measured in geometries that eliminate the
need for  complicating additional components such as the electrical wires
of the thermometer itself.
 Differential measurements
allow easy monitoring of temperature drifts in the cryogenic environment.
The technique is rapid to use and easily calibrated. 
Preliminary results will be discussed.

\end{abstract}


\section{Introduction}

The problem of characterizing the thermal properties of mesoscopic thin-film structures 
at low temperatures (below $3\,{\rm K}$), in
particular their thermal conductances and heat capacities as a function of temperature,  remains
one of the key challenges for designers of ultra-low-noise detectors. 
This problem is not confined exclusively to the detector community.\cite{giazotto_review}
These measurements require small, easily fabricated, easily
characterized thermometers. Techniques are already used
 such as Johnson noise thermometry (JNT) using
thin film resistors as noise sources with dc-SQUID readout or 
measurements of thermal properties using Transition Edge Sensors (TESs) but both have practical limitations.  JNT can perform measurements over a reasonable
temperature range and is in principal a primary thermometer 
but is in practice  secondary because of  stray resistance in the input circuit to the SQUID
 that must be calibrated. 
The achievable measurement precision,$\sigma_T$, for a source at temperature $T_s$ is given by the radiometer equation
\begin{equation}
\sigma_T^2=\frac{T_s^2}{t_m \Delta f  } 
\label{Eq:radiometer}
\end{equation}
where $t_m$ is the measurement time and $\Delta f$ is the measurement bandwidth.\cite{Kraus_radio,White_review}
 In practice the bandwidth is limited
by the source resistance and the input inductance of the SQUID to a few 10's of kHz. This gives $\sigma_T\simeq3\, {\rm mK}$ for
$T_s=500\,{\rm mK}$ with $t_m=1\,{\rm s}$. This is the precision that we found in practice.\cite{Karwan2008}
TESs can be used to determine conductances by measuring the power dissipated in the active region of the device where
electrothermal feedback (ETF) stabilizes the TES at its transition temperature, $T_c$, as a function of the
temperature of the heat bath, $T_b$. 
These measurements are widely reported 
particularly in the context of 
measurements of the thermal conductance of silicon nitride films, but a key
limitation is that the technique measures  thermal properties averaged
 over a large temperature difference (i.e. between $T_c$ and $T_b$) and the films under
study must support additional (generally superconducting) metalization to provide electrical connection. 
A problem arises if the  thermal properties are themselves a function of temperature.
A technique
for measuring conductances or heat capacities of micron-scale objects rapidly over a reasonable temperature range 
without additional overlying films
certainly seems to be required. The proposed technique permits true differential measurements of conductance (i.e with small
temperature gradients), in a geometry without complicating additional metalization. 
Measurements of heat capacities are also possible. The technique is easy to implement, simple to 
calibrate and rapid to use.

We recently demonstrated highly efficient coupling of very broadband thermal power
between the impedance-matched termination resistors of a 
superconducting microstrip transmission line.\cite{Karwan_LoC1} 
The efficiency of a short microstrip ($l\simeq2\, {\rm mm}$  was better than 
97\% for source temperatures up to $1.5\,{\rm K}$. 
When the coupling efficiency is very high the 
power transferred along a superconducting microstrip between a source and a TES can 
be used as a fast, accurate thermometer.~\cite{Tes_thermometer}  
The practical temperature sensitivity
is determined by the TES low-frequency Noise Equivalent Power, $NEP(0)$. 
The limiting temperature sensitivity is determined by the very-wide radio-frequency bandwidth
of the microstrip coupling so that we substitute $\Delta f\to \Delta \nu$ in Eq.~\ref{Eq:radiometer}
and $\Delta \nu=34.5/T_s\,{\rm GHz/K}$ is the equivalent rf-bandwidth of
 a blackbody source at temperature $T_s$.

\section{TES thermometry} 
\label{sec:TES_thermometry}  
We begin by reviewing the technique of TES thermometry for a simple
 geometry. In section~\ref{subsec:full thermal} 
we describe the full thermal circuit used in the measurements of conductances. In sections~\ref{subsec:conductance calc}
and \ref{subsec:conductance_Gb} we describe how the conductances are determined experimentally.

\subsection{A simple geometry}
\begin{figure}[!t]
\centering
\includegraphics[width=3.9cm]{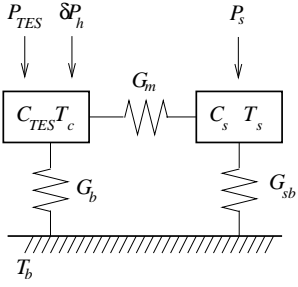}
\caption 
   { \label{fig:simple_thermal} 
Thermal circuit of a simple geometry. 
}
\end{figure}

A thermal circuit for a simple geometry for TES thermometry is shown in Fig.~\ref{fig:simple_thermal}. 
A source of total heat capacity $C_s$ which could be a thermally isolated silicon nitride (${\rm SiN_x}$) island is connected to 
a TES also formed on a ${\rm SiN_x}$ island by a 
 conductance formed from a 
resistively-terminated superconducting microstrip. 
Broadband power is transferred between the impedance-matched termination resistors of the 
microstrip which  has thermal conductance $G_m$. Over a portion of its length the microstrip crosses the Si
substrate so that any phonon conductance is efficiently heat sunk to the bath. $G_m$ arises from  conduction due to photons.
The source and TES are
connected to a heat bath at temperature $T_b$ by conductances $G_s$ and $G_{sb}$ respectively which include the
contribution from the phonon conductance of dielectric of the microstrip.
Heaters permit the source temperature to be
varied and the TES to be calibrated. 
%
%
Changes in the power coupling along the microstrip are measured using the TES which is  
in close proximity 
to its termination resistor. For low source temperatures $T_s \le  3\,{\rm K}$, all of the power is contained within the
pair-breaking threshold of the superconducting Nb, $2\Delta_{Nb} /h\simeq 760\,{\rm GHz}$ where
$h$ is Planck's constant and $\Delta_{Nb}$ is the superconducting energy gap. The power transmitted
between the source at temperature $T_s$  and the TES at its transition temperature $T_c$ is given by
\begin{equation}
P_m(T_s,T_c)= \int_0^ { \frac {2\Delta_{Nb} }{h}} [P_\nu(T_s)-P_\nu(T_c)] d\nu , 
\label{Eq:P_microstrip}
\end{equation}
where $\nu$ is the frequency, $P_\nu(T_i) =  h\nu\, n(\nu,T_i)$ 
 and  $n(\nu,T_i)$ is the Bose-Einstein distribution, and 
we have assumed that the coupling is loss-less. If the source temperature is low $T_s\le 3\,{\rm K}$
all of the power is contained well-within the cut-off frequency of the microstrip
and the upper limit of 
may be set to infinity.
Eq.~\ref{Eq:P_microstrip} then has the solution
\begin{equation}
P_m(T_s,T_c)= \frac {\pi^2 k_b^2}{6 h} (T_s^2-T_c^2) .
\label{Eq:P_microstrip_analytic}
\end{equation}
The measurement conductance $G_m(T_s)= dP_m(T_s,T_c)/dT_s$ is
 $G_m(T_s)=\pi^2 k_b^2 T_s/3h$. $G_m$ determines how changes in source temperature affect power flow to the TES.

To calibrate the thermometry we need to know the low-frequency TES current-to-power responsivity, $s_I(0)$. This
 is measured by applying slowly-varying power, $\delta P_{h}$, to the known heater resistance on 
the TES island and measuring the change in detected current $\delta I$. The responsivity
is then $s_I(0)= \delta I/ \delta P_{h}$. 
As the source temperature is changed using the source heater, the power incident on the TES changes
as given by Eq.~\ref{Eq:P_microstrip_analytic}.
Measuring the change in the current flowing through the TES, $\delta I$,  
determines the change in detected
power $\delta P_{m}= \delta I /s_I(0)$. 
Noting that $\delta P_{m}= P(T_s, T_c)-P(T_b,T_c)$,  
and since ETF fixed the  TES's temperature at  $T_c$, we can determine the source temperature 
to a good approximation as
\begin{equation}
T_s=\sqrt{  \frac{\delta I}{s_I(0)} \frac{6 h}{\pi^2 k_b^2} + T_{b}^2   } .
\label{Eq:How_calc_Ts}
\end{equation}
The practical temperature measurement precision is determined by the low-frequency TES $NEP(0)$ 
so that\cite{Tes_thermometer} 
\begin{equation}
\sigma_{T_s}^2 =   \frac{NEP^2(0)} {2 t_m G_m^2(T_s)}   .
\label{Eq:sigma_T_prac}
\end{equation}
Here we have intentionally omitted thermal fluctuations of the source island which we
 consider a signal in this geometry.
For  $NEP(0)=2\times10^{-17}\,{\rm W/\sqrt{Hz}}$ the achievable temperature precision is $30\,{\rm \mu K}$
for $t_m=1\,{\rm s}$ and $T_s=500\,{\rm mK}$. This is two orders of magnitude
better than JNT.
\subsection{The measured geometry}
\label{subsec:full thermal}
\begin{figure}[!t]
\centering
\includegraphics[width=7.5cm]{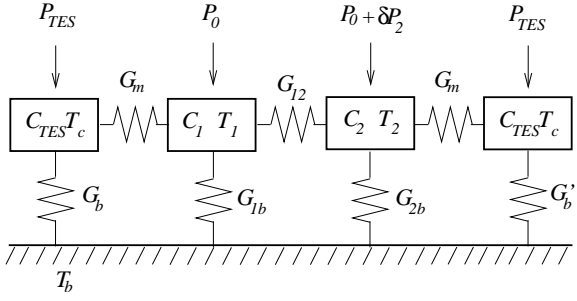}
\caption 
   { \label{fig:full_thermal} 
Thermal circuit of the full measurement. The subject under test is shown here as a conductance
$G_{12}$ but may be more complicated.
}
\end{figure}

In the full geometry  two source islands $S_1$, $S_2$ are connected by
the subject under test here a conductance, $G_{12}$. The subject may be more complicated.
Each source island is connected to its own TES by a microstrip. Figure~\ref{fig:full_thermal} 
shows the full thermal circuit.
%
We measure the quiescent TES currents to monitor and subtract
small drifts in the substrate temperature or the electronics. 
The measurement precision  of the temperature of either source, $\sigma_T$, with a measurement time $t_m$
 is determined by the low-frequency TES $NEP(0)$
 so that
\begin{equation}
\sigma_{T_{1,2}}^2  = \frac{NEP^2(0)}{2t_m} \left(\frac {1}{G_m^2(T_b)}+ \frac{1}{G_m^2(T_{1,2})}\right) 
+\frac {k_b T_{1,2}^2}{C_{1,2}}
\label{Eq:sigma_T},
\end{equation}
and the temperature measurement precision for this  differencing approach includes a contribution
from the bath temperature measurement. We have also now explicitly
included  the effect of thermodynamic
fluctuations in the temperatures of the sources of heat capacity $C_{1,2}$ since these fluctuations
directly affect the precision with which the source temperature can be determined.

\subsection{Measurement of the conductance $G_{12}$}
\label{subsec:conductance calc}
\begin{figure}[!t]
\centering
\includegraphics[height=7.5cm, angle=-90]{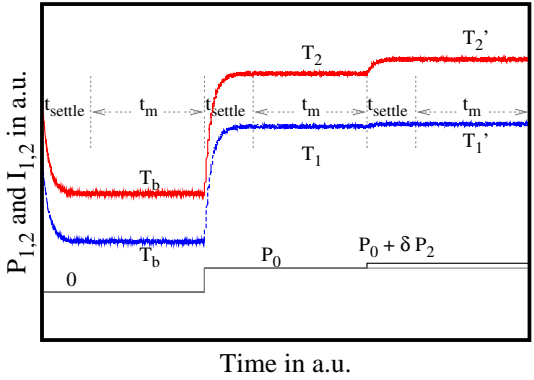}
\caption 
   { \label{fig:Measurement_cycle} 
Schematic showing the time variation of the input powers (lower traces) and output currents (upper traces)
for one cycle of the measurement. 
The sequence of measurement and settling times is indicated. The full time-sequence
represents a measurement frame. The measurement intervals determine the temperatures as shown.
}
\end{figure}

We measure the interconnect conductance $G_{12}$ in a three step process. 
A schematic of the measurement cycle is shown in Fig.~\ref{fig:Measurement_cycle}.
In the first step 
the quiescent TES currents are measured. Then
dc power $P_0$ is input to both source islands raising the temperatures from the bath temperature
$T_b$ to $T_1$, $T_2$.   
We find $T_1 \cong T_2$ and we quantify the effect of $T_1 \ne T_2$ later.
 Finally  power to $S_2$ is  stepped by an additional small amount $\delta P_2$ 
raising the island temperatures to $T_1^\prime$ and 
$T_2^\prime$. 
Since both $T_1^\prime -T_1$ and $T_2^\prime -T_2$ are small the measurement is differential.
For the next temperature sample the process is repeated with incremented $P_0$. The
power steps are adjusted in software to give approximately constant increments and differences in temperature 
 across the range of temperatures measured.
One complete measurement cycle (i.e. $P=0, P_0, P_0+\delta P_2$), with a short settling
time $t_{settle}$ between each measurement step to accommodate thermal response times, defines a `frame' time. The
 frame must be measured in a time less than the
 Allan time of the system.

The power flow  across a conductance $G$ connecting thermal reservoirs at 
temperatures $T$, $T^\prime$ can be written for notational convenience as
\begin{equation}
P(T^\prime,T) = \int_{T} ^ {T^\prime} G(T) dT = \overline{G(T^\prime,T)} (T^\prime - T)  , 
\label{Eq:P_bath}
\end{equation}
where the over-set line denotes averaging. If $ T^\prime - T = \delta T $ is small then the power flow
can be linearized so that
$P(T^\prime,T) = G(\overline{T}) \delta T$.  
Ignoring for now the small conductance of the microstrips, the input powers and resultant temperatures are related by
\begin{subequations}
\begin{align}
P_0 & =  P_{1b}(T_1,T_b) - \overline{G_{12}(T_1,T_2)} (T_2-T_1) \label{Eq:P_bath_0_a} \\
P_0 & =  P_{2b}(T_2,T_b) + \overline{G_{12}(T_1,T_2)} (T_2-T_1) \label{Eq:P_bath_0_b} \\
P_0 & =  P_{1b}(T_1^\prime  ,T_b) - \overline{G_{12}(T_1^\prime,T_2^\prime)}(T_2^\prime-T_1^\prime) \label{Eq:P_bath_0_c} \\
P_0 + \delta P_2 & =  P_{2b}(T_2^\prime  ,T_b)  + \overline{ G_{12}(T_1^\prime,T_2^\prime)}(T_2^\prime-T_1^\prime)
\label{Eq:P_bath_0_d}
\end{align}
\end{subequations}
Subtracting \ref{Eq:P_bath_0_a} from \ref{Eq:P_bath_0_c}, \ref{Eq:P_bath_0_b} from \ref{Eq:P_bath_0_d} and
using, for example, 
\begin{equation}
\begin{split}
P_{2b}(T_2^\prime  ,T_b) - P_{2b}(T_2  ,T_b))  & = \int_{T_2}^{T_2^\prime} G_{2b}(T)dT \\ 
& =G_{2b}(\overline {T_2} )\delta T_2, 
\label{Eq:Example_gbar}
\end{split}
\end{equation} 
where $\overline{T_2}= (T_2^\prime+T_2) /2$ and the final equality follows since
 $\delta T_2$ is small,  we find
\begin{equation}
\delta P_2 = 2 G_{12}(\overline{ T_{12}})(\delta T_2 - \delta T_1)  + G_{2b}(\overline{T_2})\delta T_2 - G_{1b}(\overline{T_1}) \delta T_1
\label{Gs_determined}
\end{equation}
and $\overline{T_{12}}= (T_1+T_1^\prime+T_2+T_2^\prime)/4$. 
Finally the effect of conductance along the microstrip needs to be included. The result is
\begin{multline}
 G_{12}(\overline {T_{12}}) = \\
 \frac { \delta P_2 - \left( G_{2b}(\overline{T_2})+ G_m(\overline{T_2})\right) \delta T_2 + 
\left( G_{1b}(\overline{T_1})+ G_m(\overline{T_1})\right) \delta T_1  }   
 {2(\delta T_2 - \delta T_1)} \\
   .
\label{Gs_determined_full}
\end{multline}

\subsection{Measurement of the conductances $G_{1b}$, $G_{2b}$}
\label{subsec:conductance_Gb}
The conductance to the bath at a given temperature is measured in a two-step procedure where we measure
the quiescent  TES current ($P=0$) and the effect of  applying equal power $P_0$ to both 
islands and measuring the changes in $T_1$ and $T_2$. The next
temperature sample uses incremented $P_0$. This measurement is rapid. With $t_m=0.82\,{\rm s}$,  500 data
points are acquired in less than 15 minutes.  
In the analysis we use Eqs.~\ref{Eq:P_bath_0_a} and \ref{Eq:P_bath_0_b} and assume $T_1=T_2$. This introduces an inevitable error in the 
analysis and the magnitude is of order 
$\epsilon=\vert G_{12}(\overline{T_{12}})(T_2-T_1)\vert / P_{1,2b}$. Experimentally
the error is small $\ll 1\% $. 
A fifth-order polynomial is fitted to the $T_{1,2}-P_0$ data and the conductances $G_{1b}$, $G_{2b}$ found by differentiation.
Note that since the temperature difference $T_{1,2}-T_b$ may be large this measurement is temperature-averaged.
\section{Measured device}
\label{sec:Device_description}
\begin{figure}[!t]
\centering
\includegraphics[width=7.5cm]{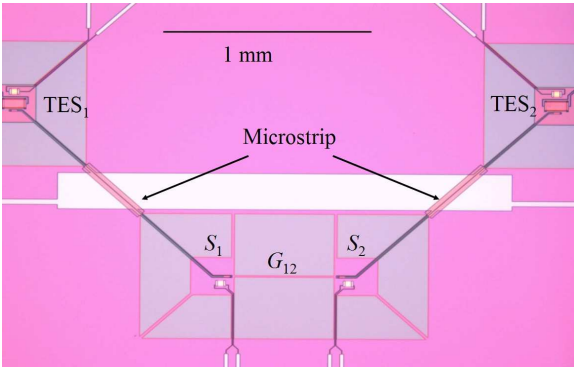}
\caption 
   { \label{fig:photo} 
Optical image of the device under test. The central nitride bar (labelled $G_{12}$) of length $500\, {\rm \mu m}$, width $10\, {\rm \mu m}$ connects
the nitride source islands $S_1$, $S_2$. Resistively terminated microstrip lines run from the islands to TESs in the upper corners of the image. The conductance 
between each nitride island and the heat bath is formed from four nitride legs, two of length $285\, {\rm \mu m}$, two of $355\, {\rm \mu m}$
that run diagonally  . 
}
\end{figure}

An optical image of the device described here is shown in Fig.~\ref{fig:photo}. 
The measured conductance 
labelled $G_{12}$ is a long thin ${\rm SiN_x}$ bar of dimensions 
$500\times 10\times 0.5 \, {\rm \mu m^3}$ formed by reactive ion and deep reactive ion etching
of a nitride-coated Si wafer. $G_{12}$ carries no additional films. $G_{12}$ connects 
two larger $0.5\,{\rm \mu m}$-thick nitride source  islands $S_1$ and $S_2$
 themselves
isolated from the Si wafer by four supporting nitride legs, two of length $255\,{\rm \mu m}$ two 
of length $358\,{\rm \mu m}$ each of width $15 \,{\rm \mu m}$. One of the longer legs also carries a Nb microstrip line. 
The microstrip is terminated by an impedance-matched AuCu termination resistor at each end. Each
island also supports a square AuCu resistor with Nb bias lines that can be used as a heater to modulate the 
temperature of the sources. The microstrips run  over the Si wafer then onto nitride islands which support
TESs. Routing of the microstrips in this way
ensures that the phonon conductance associated with the microstrip dielectric
is efficiently heat-sunk to the bath. 
The TES islands also include AuCu resistors with Nb bias lines that allow the power-to-current responsivity
of the TESs to be measured. 

AuCu resistors are $40\,{\rm nm}$ thick, Nb bias lines and the ground plane of the microstrip
are $250\,{\rm nm}$ thick. The microstrip dielectric is sputtered ${\rm Si0_2}$ 
and is $400\,{\rm nm}$ thick. The TESs use our standard higher temperature MoCu bilayer layup with
$40$ and $30\,{\rm nm}$ of Mo and Cu respectively. The fabrication route is identical with our
usual process for MoCu TESs.\cite{Dorota2008}
The TESs are voltage biased and read-out with SQUIDs.
The device is measured in a He-3 refrigerator with a base temperature of 259~mK.
  The transition temperature of the TESs described here was $T_c\sim 485\,{\rm mK}$ which is slightly
higher than reported in our earlier work with the same Mo-Cu bilayer lay-up. As a result the conductance
to the bath of the TESs is increased and the measured TES Noise Equivalent Power is
 $NEP(0) = 1.2\times 10^{-16}\, {\rm W Hz^{-1/2}}$.

\section{Results and Discussion}

%
%
\begin{figure}[!t]
\centering
\includegraphics[height=7.5cm, angle=-90 ]{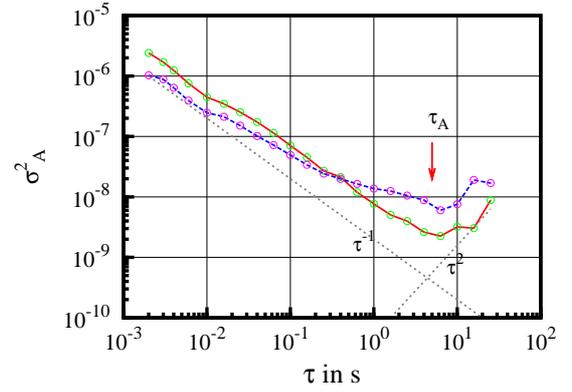}
\caption[JAPthermal] 
   { \label{fig:Allan} 
Allan plot for the two SQUID systems with biased TESs and $T_b=259\pm0.5\,{\rm mK}$. The optimum integration time,
identified as $\tau_A$, is at the minimum of the statistic $\sigma^2_A$.}
\end{figure}
The Allan variance statistic, $\sigma_A^2(\tau)$ where $\tau$ is the integration time,
provides an exceptionally powerful diagnostic of  system
stability and a plot of the variance as a function 
of  $\tau$ identifies the optimum time for
signal averaging.\cite{Allan_1966,Schieder_2001}
 In the Allan plot, 
a log-log
plot of $\sigma_A^2$ as a function of $\tau$,  
 underlying fluctuations
with frequency-domain power spectra varying as $1/f^\alpha$ show a $\tau^{(\alpha-1)}$ dependence. 
Hence
white noise with $\alpha=0$ has a $\tau^{-1}$ characteristic. $1/f$ noise shows no dependence on integration time and
drift exhibits  a dependence with $1<\alpha<3$.
Figure~\ref{fig:Allan} shows measured Allan variances for both SQUID systems with biased TESs
and the bath temperature at $T_b=259\pm0.5\,{\rm mK}$.  
The measured characteristic indicates that white noise  is reduced by time-integration up to a maximum 
of order $\tau_A\simeq 5\,{\rm s}$. 
The dashed lines show the expected behaviour for white noise with a slope $\tau^{-1}$
 and for drift
with a slope of $\tau^{2}$. This shows that drift  limits these measurements.
Guided by the Allan plot,
we chose a sample time $t_m=0.82\,{\rm s}$ being $2^{14}$ data points sampled at 
$20\,{\rm kHz}$ and  $t_{settle}=100\,{\rm ms}$. For the conductance measurements with three
power steps the total frame time is 2.8 s at each sample temperature.

%
%
\begin{figure}[!t]
\centering
\includegraphics[height=7.5cm, angle=-90 ]{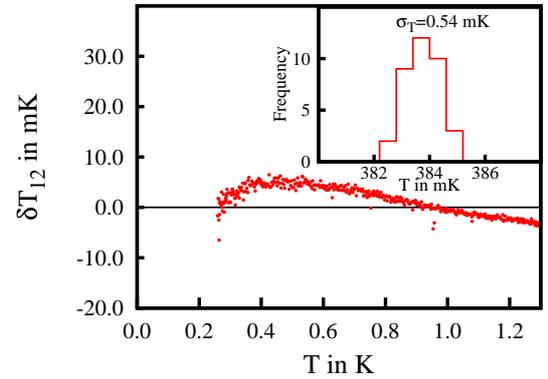}
\caption[dT] 
   { \label{fig:T_diff} 
Difference in temperature between the two islands as a function of average temperature for equal applied powers. Note the improvement in the measurement
precision as the temperature increases. The inset shows the measured temperature precision for repeated 
measurements near 384~mK. The 
measured precision $\sigma_T=0.54\, {\rm mK}$ compares well with the calculated value from Eq.~\ref{Eq:sigma_T} which
gives $\sigma_T=0.47\, {\rm mK}$ for the measurement time used $t_m=0.82\,{\rm s}$ and the measured 
$NEP(0)=1.2\times 10^{-16}\,{\rm WHz^{-1/2}}$.
}
\end{figure}
Applying equal power to both sources,  $G_{1b}$ and $G_{2b}$ were determined.
As discussed earlier the analysis 
introduces an unavoidable error in the measurement of $G_{1b}$ and $G_{2b}$ depending on $T_2-T_1$. 
Figure~\ref{fig:T_diff}
shows the difference in temperature $\delta T_{12}= T_2-T_1$ for equal powers applied to both islands as a 
function of average island temperature. 
The difference is small and a maximum of about $5\,{\rm mK}$ at 500~mK. 
Since  $\delta T_{12}$ is small we will see that the error is negligible. 
The 
temperature difference implies a difference in conductance for the two notionally
identical conductances $G_{1b}$, $G_{2b}$ of about $\pm2\%$ between 0.26 and 1.3~K. The temperature dependence 
of $\delta T_{12}$ evident in Fig.~\ref{fig:T_diff}
is also unexpected. It does not seem that the difference can be accounted for by
experimental uncertainty (such as in the calibration of the TES responsivities). One possibility
might be differences in the actual coupling efficiencies of the microstrip lines.

The inset of Fig.~\ref{fig:T_diff}
shows a histogram of calculated temperatures for repeated measurements of a source temperature near $384\,{\rm mK}$.  
The 
measured precision $\sigma_T=0.54\, {\rm mK}$ compares well with the value calculated from Eq.~\ref{Eq:sigma_T} which
gives $\sigma_t=0.47\, {\rm mK}$ for the measurement time used $t_m=0.82\,{\rm s}$ and the
measured $NEP(0)=1.2\times 10^{-16}\,{\rm WHz^{-1/2}}$.

%
%
\begin{figure}[!t]
\centering
\includegraphics[height=7.5cm, angle=-90 ]{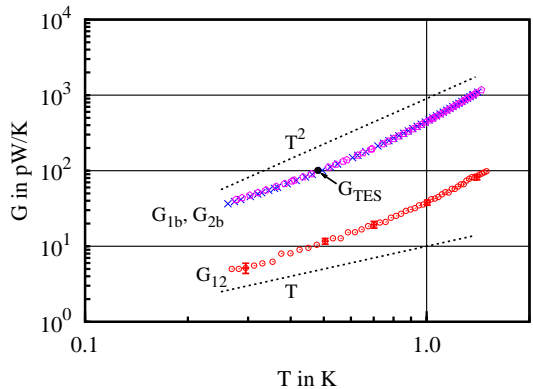}
\caption[Fig3] 
   { \label{fig:Conductances} 
Measured conductances to the bath $G_{1b}$, $G_{2b}$ and calculated sample conductance $G_{12}$. 
Estimated errors for $G_{12}$ are indicated. The dotted lines indicate  dependencies proportional to 
$T^2$ and $T$. The single circle is the measured $G_{TES}$ scaled by the $A/l$ ratios to compare directly 
to
$G_{1b}$ and $G_{2b}$.
}
\end{figure}
Figure~\ref{fig:Conductances}
shows measured conductances  $G_{1b}$, $G_{2b}$ 
and $G_{12}$ as a function of temperature. Representative
error bars for $G_{12}$ determined from the variance of repeated measurements are indicated.
We can now estimate the magnitude of the error in $G_{1b}$ and $G_{2b}$. The maximum temperature 
difference  occurs near $500\,{\rm mK}$
where $\delta T_{12}\simeq 5\, {\rm mK}$ is greatest. We find $\epsilon \simeq 0.2\%$, but less than this over most of the
temperature range. This is considered acceptable.
We also show the variation with temperature if $G=kT^\beta$ with $k$ a constant. The dotted lines show dependencies $\beta=1$ and 2.
 There is a strong suggestion here that a simple power
law does not account for the conductance across this measurement range. At the highest temperatures $\beta>2$, at the lowest 
$\beta<2$. A reduction
of the exponent may be expected at low temperatures if dominant phonon wavelengths start
to become comparable to the nitride thickness. The single dot in Fig.~\ref{fig:Conductances}
is the measured $G_{TES}$ for one of the TESs obtained in the standard way by 
measuring the power plateau in the
ETF region of the current-voltage characteristic as a function of the bath temperature. 
This single point represents approximately 2 hours of data taking. By contrast the conductances to the bath
$G_{1b}$, $G_{2b}$ plotted in Fig.~\ref{fig:Conductances}
are found from 500 measurements of temperature acquired in 15~minutes. A comparable acquistion time
measures $G_{12}$ with 50 data points with additional signal averaging.


%
%
\begin{figure}[!t]
\centering
\includegraphics[height=7.5cm, angle=-90 ]{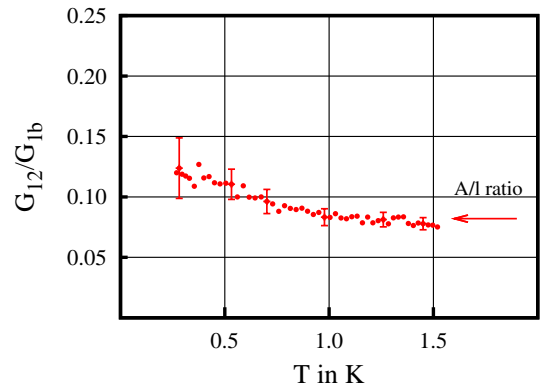}
\caption[Fig4] 
   { \label{fig:Gs_ratio} 
Ratio of the measured conductance $G_{12}$  to the conductance to the bath $G_{1b}$. The arrow indicates the expected
ratio if the conductances scale as the area to length ratio of the structures.
}
\end{figure}
Figure~\ref{fig:Gs_ratio}
shows the ratio of $G_{12}$ to $G_{1b}$ as a function of temperature.
The simplest model for power flow along, or the conductance of, a uniform bar would assume that the
power scales as the ratio of cross-sectional area, $A$, to length, $l$.
 From the dimensions of
the nitride support legs for the source islands and the  dimensions of $G_{12}$ we calculate a
conductance ratio of 0.082. This ratio includes the conductance associated with the ${\rm Si0_2}$ dielectric 
of the microstrip lines and the smaller contribution from the Nb wiring all of which we assume
are equal to that of ${\rm SiN_x}$. The variation of the ratio is experimentally significant and may
already illustrate the difference between  thermal properties 
averaged over a temperature range and the true differential
measurement.

\section{Conclusion}
We have described our first true-differential measurements of thermal conductances
 of a micron-scaled object at low temperatures
using microstrip-coupled TES thermometry. The
temperature precision is already significantly greater than that achievable with JNT with the same
measurement time. 
The measurements are rapid and easily calibrated. The achieved precision already strongly
suggests that the thermal transport characteristics of the nitride structure are not
described by a simple power-law across the temperature range $0.26$ to $1.5\,{\rm K}$.
The device under test can be fabricated without additional  metalization for wiring. It should be
straight-forward to include specific layers on the nitride test structure to measure
particular thermal properties in a controlled manner. In the future we expect to be able to explore the
temperature dependence of heat capacities of 
thin films such as ${\rm SiO_2}$ including possibly measurements of time dependent heat capacity. 
We will explore the effect on conductance of superposed layers: for example, does thermal conductance
in thin multilayers really scale as total thickness?
We also see the  possibility of exploring the 
engineering of the nitride to realise phononic structures, to achieve  reductions of the conductance 
in compact nitride structures needed for the next generation of ultra-low-noise detectors.

\section*{Acknowledgment}

We would like to thank Michael Crane for assistance with cleanroom processing, David Sawford for 
software and electrical engineering and Dennis Molloy for mechanical engineering.




\bibliography{TESreferences}   

\bibliographystyle{IEEEtran}

\end{document}